\documentclass[pre,showpacs,twocolumn,bibnotes,footinbib]{revtex4}
\usepackage{bm}
\usepackage{graphicx}
\usepackage{amsmath}
\usepackage{amssymb}

\newcommand{\vare}{\varepsilon }
\newcommand{\md}{\mathrm{d}}
\newcommand{\eqb}{\begin{equation}}
\newcommand{\eqe}{\end{equation}}

\begin{document}

\title{Zener tunneling in two-dimensional photonic lattices}

\author{Valery S. Shchesnovich$^1$, Solange B. Cavalcanti$^1$,
Jandir M. Hickmann$^1$, and Yuri S. Kivshar$^2$}

 \affiliation{$^1$Instituto de F\'{\i}sica -
Universidade Federal de Alagoas, Macei\'o AL 57072-970, Brazil\\
$^2$Nonlinear Physics Center and Centre for Ultra-high bandwidth Devices for
Optical Systems (CUDOS), Research School of Physical Sciences and Engineering,
Australian National University, Canberra ACT 0200, Australia}

\begin{abstract}
We discuss the interband light tunneling in a two-dimensional periodic photonic
structure, as was studied recently in experiments for optically-induced photonic
lattices [H. Trompeter et al., Phys. Rev. Lett. \textbf{96}, 053903 (2006)]. We
identify the Zener tunneling regime at the crossing of two Bloch bands, which
occurs in a generic case of the Bragg reflection when the Bloch index crosses the
edge of the irreducible Brillouin zone. Similarly, the higher-order Zener tunneling
involves four Bloch bands when the Bloch index passes through a high-symmetry point
on the edge of the Brillouin zone. We derive simple analytical models that describe
the tunneling effect, and calculate the corresponding tunneling probabilities.
\end{abstract}


\maketitle

\section{Introduction}

Zener tunneling in tilted periodic potentials is an intriguing physical phenomenon
which occurs when the energy difference imposed on the period by a linear potential
becomes of the order of the energy gap between two nearest Bloch bands~\cite{Zen}.
A simple model  studied independently by Zener~\cite{Zen}, Landau~\cite{Lan} and
Majorana~\cite{Maj} captures the essence of this phenomenon. The well-known
physical examples include electrical breakdown in Zener diodes~\cite{ZenerD},
electrical conduction in nanotubes~\cite{nanotb} and superlattices~\cite{suplatt},
pair tunneling in Josephson junctions~\cite{Josjunct}, tunnelling of the
Bose-Einstein condensate in optical lattices~\cite{LZBEC1,LZBEC2} and an optical
analog of tunneling in waveguide arrays and photonic
crystals~\cite{PHOT1D1,PHOT2D}.

Recent observation of Zener tunneling in two-dimensional photonic
lattices~\cite{PHOT2D} calls for a nontrivial generalization of the
Landau-Zener-Majorana system, since the latter describes only the tunneling at an
avoided crossing of two Bloch bands. In two dimensions the situation of more than
two energy levels with comparable energy gaps between them is generally
unavoidable. This is due to the fact that in higher dimensions there exist more
than one distinct (i.e., not equivalent) Bragg reflection planes. Thus, at least in
some cases a $n$-level system with $n>2$ must be invoked to describe Zener
tunneling in a two-dimensional lattice. In this paper we analytically study all
possible scenarios of Zener tunneling in two dimensional square lattices. To this
end we invoke the so-called shallow lattice approximation (see the definition
below). Our results can be applied to the tunneling in two-dimensional photonic
lattices observed in Ref.~\cite{PHOT2D}.

Bloch oscillations and Zener tunneling in  two-dimensional lattices  have been
studied numerically in Refs.~\cite{Bloch2D1,Bloch2D2}. However, a complete analysis
of all possible cases was not presented. In addition,  nonlinear effects of Zener
tunneling of Bose-Einstein condensate in two-dimensional optical lattices has been
considered recently~\cite{Kon} where it was shown that the modulational instability
of Bloch waves results in an asymmetry of the resonant upper-to-lower band vs.
lower-to-upper band tunneling (see also Refs.~\cite{WN,ZG,PHYSB} for other
nonlinear effects of tunneling in one-dimensional lattices). In our case,
nonlinearity is negligible and we restrict ourselves to the linear Zener tunneling.

The paper is organized as follows. In section \ref{sec2} we consider the general
model for the beam propagation in a photonic lattice and discuss the shallow
lattice approximation. Section \ref{sec3} contains the derivation of the
Landau-Zener-Majorana type models for Zener tunneling in a shallow lattice.
Conclusions and perspectives are summarized in section \ref{sec4}.

\section{Model and shallow-lattice approximation}
\label{sec2}
We describe the propagation of an optical beam by the paraxial equation for the
normalized electric field envelope $E$
\eqb i \frac{\partial E}{\partial \zeta}
+\frac{1}{2}\left(\frac{\partial^2 E}{\partial \xi_1^2} +
\frac{\partial^2 E}{\partial \xi_2^2} \right) + \varkappa\Delta
n(\xi)E = 0, \label{EQ1}\eqe
where $\bm{\xi} = (\xi_1,\xi_2)$ are dimensionless transverse
coordinates: $\xi_1 = (2\pi/d)x$, $\xi_2 = (2\pi/d)y$; $\zeta =
2\pi\lambda/(n_0d^2)z$ is the dimensionless propagation distance,
with $\lambda$ being the wavelength in vacuum, $n_0$ the bulk
refractive index, and $d$ the lattice period; $\varkappa =\gamma
n_0(d/\lambda)^2$ with $\gamma$ being the nonlinear coefficient.
The total induced refractive index pattern $\Delta n(\bm{\xi})$ is
given as follows
\eqb \Delta n = \frac{I_g(\bm{\xi})+I_m(\bm{\xi})}{1
+I_g(\bm{\xi})+I_m(\bm{\xi})}, \label{EQ2}\eqe
where the potential
\eqb
I_g(\bm{\xi}) = A^2\cos^2\left(\frac{\xi_1}{2}\right)
\cos^2\left(\frac{\xi_2}{2}\right)
\label{POT}\eqe
is due to an optical lattice and the additional potential
\[
I_m(\bm{\xi}) = \frac{B}{2}\left[1 +\tanh(\kappa\bm{e}_R\bm{\xi})\right],
\]
where $\kappa = {d}/({2\pi\eta})$,  describes the the refractive index ramp in the
direction $\bm{e}_R$ (the notations correspond to those of Ref.~\cite{PHOT2D}).

A shallow lattice approximation corresponds to the condition
$\varkappa \Delta n \ll 1$ [see Eq.~(\ref{EQ1})]. For instance, in
the experiment reported in Ref.~\cite{PHOT2D} the induced optical
lattice had $\varkappa =21.59$. Thus to have a weak lattice case
within the same experimental parameters one should use a weak cw
laser beam satisfying the condition $A^2\ll 1/{\varkappa}\approx
0.05$.

Assuming that a weak ramp field  ($B\ll 1$) changes slowly over
the lattice period, i.e. $2\pi\kappa B= Bd/\eta\ll A^2$, and
therefore the Bloch band structure is
preserved~\cite{note,Wan,WS}, we obtain in this case
\eqb \varkappa\Delta n = \varkappa I_g(\bm{\xi}) +\frac{1}{2} \varkappa B +
\bm{\alpha}\bm{\xi}, \label{EQ3} \eqe
where $\bm{\alpha} = \varkappa B\kappa\bm{e}_R/2 = \gamma
n_0Bd^3/(4\pi\lambda^2\eta)\bm{e}_R$. Further discussion of validity of the
Landau-Zener-Majorana  models derived below is given in the next section.

Instead of $\varkappa I_g(\bm{\xi})$,  we will use the combined potential
\eqb
V = V_0[\cos(\xi_1) +\cos(\xi_2) + \vare\cos(\xi_1)\cos(\xi_2)]
\label{lattpot}\eqe
where $V_0 = \varkappa A^2/4$, and $\vare$ is a parameter which enables us to
establish the effect of the non-separability of the photonic lattice
$I_g(\bm{\xi})$ on the tunneling probability.  We also drop the insignificant
constant terms $\varkappa B/2$ and $\varkappa A^2/4$. The weak ramp condition now
reads $|\bm{\alpha}|\ll V_0$.

\begin{figure}[ht]
\begin{center}
\includegraphics{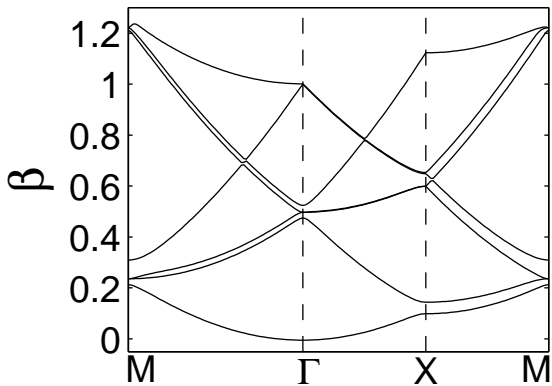}
\caption{\label{FGbands} Bloch band structure (the first 7 bands) corresponding to
the lattice of Eq. (\ref{lattpot}). Here $V_0=0.05$ and $\vare = 1$. }
\end{center}
\end{figure}

The lattice (\ref{POT}), also called the ``quantum antidot'' in the field of
Bose-Einstein condensates \cite{Bloch2D2}, corresponds to $\vare = 1$. Its band
structure for $V_0 = 0.05$ is shown in figure \ref{FGbands}.

\section{Simple models for Zener tunnelling}
\label{sec3}
An optical beam propagating in the lattice experiences a strong reflection at the
resonant Bragg planes (lines in two dimensions), i.e. the Bragg planes which are
defined by the reciprocal lattice vectors for which the lattice potential has
non-zero Fourier components (see, e.g., Refs.~\cite{Ziman,AshMer}). Let us first
briefly recall the basic theory. In the Fourier space the equation for the periodic
Bloch wave $\varphi_{\bm{q}}(\bm{\xi})$, i.e. $E =
\exp\{i\beta\zeta+i\bm{q}\bm{\xi}\}\varphi_{\bm{q}}(\bm{\xi})$,  reads
\eqb
\left[(\bm{q-\bm{Q})}^2 - \beta\right]{C}_{\bm{q}-\bm{Q} }  +
\sum_{\bm{Q}^\prime}\hat{V}_{\bm{Q}^\prime-\bm{Q}}{C}_{\bm{q} -
\bm{Q}^\prime}=0 \label{EQ4}\eqe
where $\bm{Q}$ denotes a vector of the reciprocal lattice and
$C_{\bm{q}-\bm{Q}}$ and $\hat{V}_{\bm{Q}^\prime-\bm{Q}}$ are the
Fourier components of the Bloch wave $\varphi_{\bm{q}}(\bm{\xi})$
and the lattice $V(\bm{\xi})$, respectively. The probe beam with
the Bloch index $\bm{q}=(k,\ell)$  is effectively reflected by the
lattice in the first order in $V_0$ when the end-point of the
Bloch index lies on the resonant Bragg plane defined by
$\bm{Q}^\prime-\bm{Q}$, i.e. when $(\bm{q}-\bm{Q})^2 =
(\bm{q}-\bm{Q}^\prime)^2 =\beta$ and
$\hat{V}_{\bm{Q}^\prime-\bm{Q}}\ne0$.

\begin{figure}[ht]
\begin{center}
\includegraphics{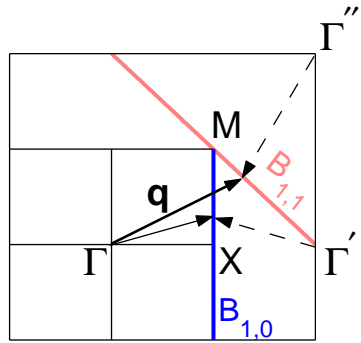}
\caption{\label{FG1} Schematic representation of the extended Brillouin zone and
the two cases of the Bragg resonance for the two-dimensional optical lattice
described in the text. Solid lines indicate the Bloch index and dashed lines are
the resonant index given by $\bm{q}-\bm{Q}$ with $\bm{Q}$ corresponding to the
considered Bragg plane. }
\end{center}
\end{figure}

For the lattice  (\ref{lattpot}) there are two distinct resonant planes, denoted by
$B_{1,0}$ and $B_{1,1}$ in Fig.~\ref{FG1}. The corresponding vectors of the
reciprocal lattice are $\bm{Q} = (2q_B,0)$ and $\bm{Q}=(2q_B,2q_B)$, respectively
(in our case, $q_B = 1/2$). Below we consider the corresponding  Bragg resonances.

\subsection{Generic $B_{1,0}$ resonance}

The reflection point on the Bragg plane $B_{1,0}$ is $\bm{q}=(q_B,\ell_0)$ with
some $\ell_0\ne \pm q_B +\mathcal{O}(V_0)$ (i.e., outside a neighborhood of the
$M$-point of radius $\propto V_0$). For a broad beam one can proceed in the way
similar to the approach of Ref.~\cite{Houst}, that studied accelerating electrons
in the solid-state theory, and assume that the beam is described by a Bloch wave
with the propagation-dependent Bloch index $\bm{q}=\bm{q}(\zeta)$. By keeping only
the resonant terms  we arrive at the following approximation for the Bloch wave
describing a broad beam near the point of Bragg reflection (or, equivalently,
resonance)
\eqb \psi
= \left( C_1(\zeta)e^{ik(\zeta)\xi_1} +
C_2(\zeta)e^{i[k(\zeta)-2q_B]\xi_1}\right)e^{i\ell(\zeta)\xi_2}.
\label{EQ5}\eqe
Substituting this expression into Eq.~(\ref{EQ1}), with $\varkappa\Delta n =
V(\bm{\xi}) +\bm{\alpha}\bm{\xi}$, and projecting on the resonant terms yields
$dk/d\zeta=-\alpha_1$, $d\ell/d\zeta=-\alpha_2$, which are necessary to cancel the
linear term in $\bm{\xi}$ (see, e.g., Ref.~\cite{PHYSB}), and then obtain a system
of coupled equations for the incident $C_1$ and Bragg reflected $C_2$ amplitudes
(see also Ref.~\cite{WN}),
\begin{eqnarray}
i\frac{dC_1}{d\zeta} &=& \frac{1}{2}(\ell^2 +k^2)C_1 +\frac{V_0}{2}C_2,\nonumber\\
& & \label{EQ6}\\
i\frac{dC_2}{d\zeta} &=&  \frac{1}{2}(\ell^2 + [k-2q_B]^2)C_2
+\frac{V_0}{2}C_1.\nonumber
\end{eqnarray}
It is convenient to set the propagation variable $\zeta$ equal zero at the
resonance point, so that $k = q_B - \alpha_1 \zeta$. Redefining the amplitudes as
follows $(C_1,C_2) = \exp\{-i(\int^\zeta\md\zeta \ell^2 + q_B^2\zeta
+\alpha_1^2\zeta^3/3)/2\}(c_1,c_2)$ we obtain {\em the Landau-Zener-Majorana
system}~\cite{Zen,Lan,Maj}
\begin{eqnarray}
i\frac{dc_1}{d\zeta} &=& -\frac{\alpha_1 \zeta}{2}c_1 +\frac{V_0}{2}c_2,\nonumber\\
& & \label{EQ7}\\
 i\frac{dc_2}{d\zeta} &=&  \frac{\alpha_1\zeta}{2}c_2
 +\frac{V_0}{2}c_1.\nonumber
\end{eqnarray}
System (\ref{EQ7}) is Hamiltonian, its  adiabatic energy levels or, equivalently,
the two Bloch bands of $\beta$, $\beta_{1,2} = \mp\sqrt{q^2 +(V_0/2)^2}$,
experience  an avoided crossing at $q=0$, where $q\equiv\alpha_1\zeta/2$ has the
meaning of the running band parameter.

The probability of tunneling $P$, defined through the final amplitude
$P=|C_j(\infty)|^2$  for the initial condition $|C_j(-\infty)|=1$, is given by the
well-known Landau-Zener formula $P = \exp \{-\pi V_0^2/(2|\alpha_1|)\}$ for both
the lower-to-upper and upper-to-lower band tunneling processes. Here $\alpha_1$ is
the component of $\bm{\alpha}$ in the direction perpendicular to the Bragg plane
$B_{1,0}$, i.e. to the $XM$-border of the irreducible Brillouin zone. Thus, this
case of the interband tunneling is quasi one-dimensional even it occurs in a
two-dimensional lattice (e.g., see Ref.~\cite{WN,ZG,PHYSB}).

\subsection{Generic $B_{1,1}$ resonance}

In this case, the reflection point on the Bragg plane $B_{1,1}$ is
$\bm{q}=(k_0,\ell_0)$ with $k_0 +\ell_0 =2q_B$ and $\ell_0\ne
q_B+\mathcal{O}(V_0)$. The resonant terms of the Bloch wave read
\eqb \psi = C_1(\zeta)e^{ik(\zeta)\xi_1 +i\ell(\zeta)\xi_2} +
C_2(\zeta)e^{i[k(\zeta)-2q_B]\xi_1 +i[\ell(\zeta)-2q_B]\xi_2}.
\label{EQ8}\eqe Setting $k = k_0 - \alpha_1\zeta$ and $\ell = \ell_0
-\alpha_2\zeta$, redefining the amplitudes as $(C_1,C_2) = \exp\{-i([k_0^2+\ell_0^2
+(\alpha_2-\alpha_1)(k_0-\ell_0)]\zeta+[\alpha_1^2+\alpha_2^2]\zeta^3/3)/2\}(c_1,c_2)$
and following all the steps described above results in the following
Landau-Zener-Majorana system
\begin{eqnarray}
i\frac{dc_1}{d\zeta} &=& -\frac{\Omega_1\zeta}{2}c_1+\frac{\vare V_0}{4}c_2,\nonumber\\
& &\label{EQ9} \\
i\frac{dc_2}{d\zeta} &=&  \frac{\Omega_1\zeta}{2}c_2 +\frac{\vare
V_0}{4}c_1,\nonumber
\end{eqnarray}
where $\Omega_1 = \alpha_1 +\alpha_2$.

The corresponding Bloch bands are $\beta_{1,2} = \mp\sqrt{q^2 +(\vare V_0/4)^2}$,
with $q\equiv\Omega_1\zeta/2$. The interband tunneling probability when crossing
the $B_{1,1}$-plane is  thus given by the formula $P = \exp\{-\pi
\vare^2V_0^2/(8|\alpha_1+\alpha_2|)\}$.

The Bragg plane $B_{1,1}$ is transformed to the $\Gamma M$-border of
the irreducible Brillouin zone by the following transformation: $k =
2q_B - k^\prime$, $\ell = \ell^\prime$. Note that the transformation
changes the coordinates as follows $\xi_1 = -\xi_1^\prime$, $\xi_2 =
\xi_2^\prime$, thus $\alpha_1 = -\alpha_1^\prime$ and $\alpha_2 =
\alpha_2^\prime$. Therefore, the probability of tunneling when
crossing the $\Gamma M$-border of the irreducible Brillouin zone
reads $P = \exp\{-\pi \vare^2V_0^2/(8|\alpha_2-\alpha_1|)\}$.

The quantity $|\alpha_2-\alpha_1|/2$ is  the component of
$\bm{\alpha}$ perpendicular to the $\Gamma M$-line (exactly
$|\alpha_2-\alpha_1|/\sqrt{2}$) multiplied by a half length of the
respective  reciprocal lattice vector, $\sqrt{2}q_B$. In this
respect, this case of tunneling is again quasi one-dimensional.

\subsection{Four-fold $B_{1,1}$ resonance}

In this more general case, the reflection point is the $M$-point.
There exist exactly four Bloch indices which are in resonance,
they correspond to the four high-symmetry points of the first
Brillouin zone equivalent to the $M$-point. This case is specific
to the two-dimensional tunneling, since it leads to the four-level
system. The resonant part of the  Bloch wave reads
\[
\psi = C_1(\zeta)e^{ik(\zeta)\xi_1 +i\ell(\zeta)\xi_2} +
C_2(\zeta)e^{i[k(\zeta)-2q_B]\xi_1 +i[\ell(\zeta)-2q_B]\xi_2}
\]
\eqb + C_3(\zeta)e^{i[k(\zeta)-2q_B]\xi_1 +i\ell(\zeta)\xi_2} +
C_4(\zeta)e^{ik(\zeta)\xi_1 +i[\ell(\zeta)-2q_B]\xi_2}.
\label{EQ10}\eqe Setting  $k = q_B - \alpha_1\zeta$ and $\ell =
q_B-\alpha_2\zeta$, defining $c_j=e^{i(q_B^2\zeta +
[\alpha_1^2+\alpha_2^2]\zeta^3/6)}C_j$ and following all the above
steps we derive the following system
\begin{eqnarray}
i\frac{dc_1}{d\zeta}  &=& -\frac{\Omega_1\zeta}{2}c_1  +\frac{\vare
V_0}{4}c_2  +\frac{V_0}{2}(c_3  +c_4 ),\label{EQ11}\\
 i\frac{dc_2}{d\zeta}  &=& \frac{\Omega_1\zeta}{2}c_2
+\frac{\vare V_0}{4}c_1  +\frac{V_0}{2}(c_3  +c_4 ),\label{EQ12}\\
 i\frac{dc_3}{d\zeta}  &=& -\frac{\Omega_2\zeta}{2}c_3
+\frac{\vare V_0}{4}c_4  +\frac{V_0}{2}(c_1  +c_2 ),\label{EQ13}\\
 i\frac{dc_4}{d\zeta}  &=& \frac{\Omega_2\zeta}{2}c_4
+\frac{\vare V_0}{4}c_3  +\frac{V_0}{2}(c_1  +c_2 ),\label{EQ14}
\end{eqnarray}
with $\Omega_1 = \alpha_1+\alpha_2$ and $ \Omega_2 = \alpha_2-\alpha_1$. The
invariance of the system with respect to rotations by $\pi/2$, which is the
invariance of the lattice itself, is evident from the corresponding transformation:
$\alpha_1\to\alpha_2$ and $\alpha_2\to-\alpha_1$.

System (\ref{EQ11})-(\ref{EQ14}) can be viewed as a coupling (by the terms with
$V_0/2$) of two cases of the Bragg reflection: one is for crossing of the
$B_{1,1}$-plane [Eqs.~(\ref{EQ11})-(\ref{EQ12})] and the other one is for crossing
of the $B_{-1,1}$-plane [Eqs.~(\ref{EQ13})-(\ref{EQ14}); the reciprocal lattice
vector is $\bm{Q} = (-2q_B,2q_B)$)].

\begin{figure}[ht]
\begin{center}
\includegraphics{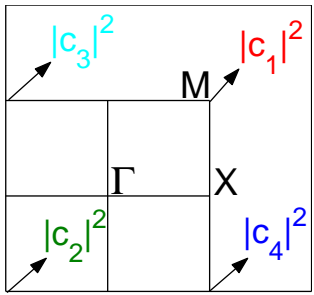}
\caption{\label{FG2} Schematic representation of tunneling at the
$M$-point of the Brillouin zone. The four beams follow the direction
of the linear ramp (indicated by arrows; here along the $\Gamma
M$-line). }
\end{center}
\end{figure}

A schematic correspondence between the amplitudes $|c_k|^2$ of the
Bloch wave and the tunneled and reflected beams is depicted in
Fig.~\ref{FG2}.

The generalized  (i.e. multilevel) Landau-Zener-Majorana system defined by
Eqs.~(\ref{EQ11})-(\ref{EQ14}) leads to a fourth-order polynomial which does not
allow a simple analytical solution for the Bloch bands. However, it is easy to
establish that the four Bloch bands assume the following values at the crossing
point ($\zeta=0$)
\eqb
\beta(M) =\frac{1}{4} + V_0\left[
\frac{\vare}{4}-1;-\frac{\vare}{4};-\frac{\vare}{4};1+\frac{\vare}{4}\right],
\label{atMpoint}\eqe
where we have used that $|\bm{q}_M| = 1/\sqrt{2}$.  Thus, recalling that in our
case $\vare=1$, we see that the shallow lattice captures at least the qualitative
behavior of the Bloch bands at the $M$-point of the experimental two-dimensional
optical lattice of Ref.~\cite{PHOT2D}, where the two inner Bloch bands are also
crossing at the $M$-point (see also figure \ref{FGbands}).

Though there exists no general formula describing the tunneling probabilities
between various levels in a multilevel Landau-Zener-Majorana system (see, e.g.,
Ref.~\cite{BE,DO,Shytov} for further discussion), in the generic case, i.e.
$\alpha_{1,2} \ne 0$, the probability of the transition between the lowest and the
highest adiabatic levels  is known~\cite{BE,Shytov}. This corresponds to the
tunneling across the complete band gap discussed in Ref.~\cite{PHOT2D}. Assuming
that $|\alpha_1+\alpha_2|>|\alpha_2-\alpha_1|$, we obtain the probability of
tunneling between the first and the fourth bands as follows
\eqb P =
\exp\left\{-\frac{\pi
V_0^2}{2}\left[\frac{\vare^2}{4|\alpha_2+\alpha_1|} +
\frac{1}{|\alpha_1|} + \frac{1}{|\alpha_2|}\right]\right\}.
\label{EQ15}\eqe
In the opposite case, i.e. when
$|\alpha_1+\alpha_2|<|\alpha_2-\alpha_1|$, the Bloch index passes
through an equivalent $M$-point and the probability of tunneling
is given by the same formula with the change
$|\alpha_1+\alpha_2|\to|\alpha_2-\alpha_1|$.

\begin{figure}[ht]
\begin{center}
\includegraphics{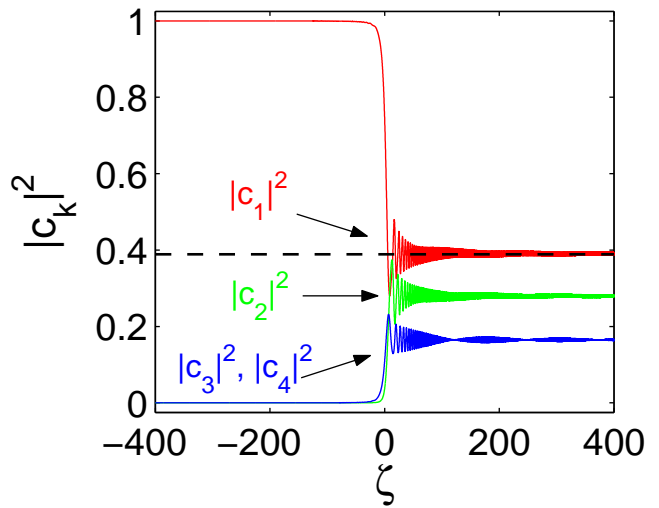}
\caption{\label{FG3} Tunneling at the $M$-point of the Brillouin
zone. Here the parameters are: $\bm{\alpha} =
0.05[\cos(\pi/4),\sin(\pi/4)]$, i.e. the ramp is in the $\Gamma
M$-direction, $V_0 = 0.1$, and $\vare=1$. Dashed line is the
analytical prediction given by the formula~(\ref{EQ15}).  }
\end{center}
\end{figure}

\begin{figure}[ht]
\begin{center}
\includegraphics{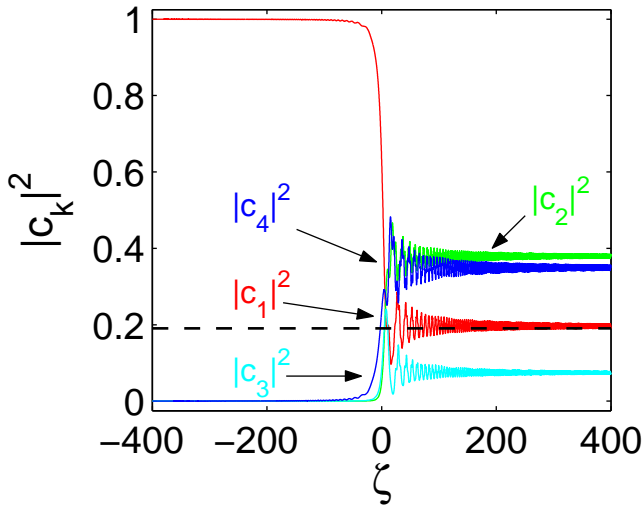}
\caption{\label{FG4} Tunneling at the $M$-point. Here the parameters
are: $V_0 = 0.1$, $\vare=1$, and $\bm{\alpha} =
0.05[\cos(0.25),\sin(0.25)]$, i.e. the ramp is directed close to the
$\Gamma X$-line. Dashed line is the analytical prediction. }
\end{center}
\end{figure}

Two particular cases of tunneling at the $M$-point  are shown in Figs.~\ref{FG3}
and \ref{FG4}. We have simulated numerically system (\ref{EQ11})-(\ref{EQ14}) using
the fourth-order Runge-Kutta method. Since the system is written for the so-called
diabatic basis, a large $\zeta$-interval was used for reliable results.
Asymptotically, i.e. for $\zeta\to\pm\infty$, the diabatic basis coincides with the
adiabatic one (in our case, the latter defines the respective Bloch bands). The
dashed line gives the analytical prediction of Eq.~(\ref{EQ15}). The oscillations
of the amplitudes $|c_k|^2$ in Figs.~\ref{FG3} and \ref{FG4} after the tunneling
process reflect a slight difference between the respective amplitudes in the
diabatic and adiabatic bases (in the adiabatic basis there are no oscillations).

In Fig.~\ref{FG3} the ramp direction is along the $\Gamma M$-line and the output
powers of the side beams (i.e. $W\propto |C_{3,4}|^2$) are the same.  In
Fig.~\ref{FG4} the ramp direction is close to the $\Gamma X$-border thus creating a
substantial asymmetry in the output beams 3 and 4, similar to the tunneling
observed in the experiment~\cite{PHOT2D}.

\begin{figure}[ht]
\begin{center}
\includegraphics{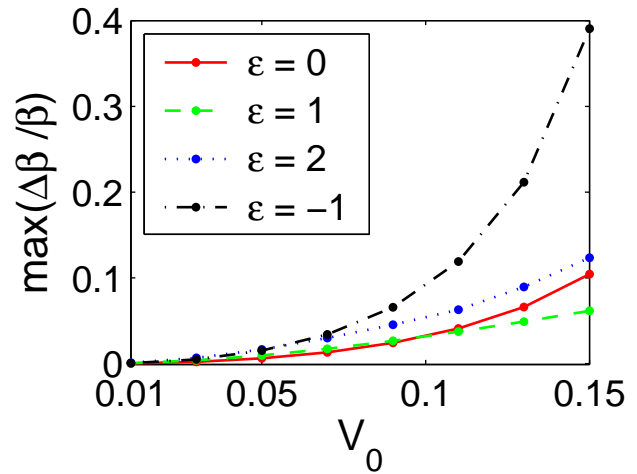}
\caption{\label{FGErr} The error of the Landau-Zener-Majorana model for the
four-fold $B_{1,1}$-resonance.  The curves give the absolute value of the maximal
relative difference between the  first four Bloch bands of the full 2D equation and
the corresponding energy levels of the four-level Landau-Zener-Majorana model.  }
\end{center}
\end{figure}

We conclude this section with discussion of validity of the Landau-Zener-Majorana
models considered above. In particular, we  consider the four-fold
$B_{1,1}$-resonance. The system (\ref{EQ11})-(\ref{EQ14}) approximates the four
resonant Bloch bands of the lattice  at the $M$-point (see figure \ref{FGbands}).
The approximate analytical values for the four energies  at the $M$-point, given by
equation (\ref{atMpoint}), can be employed to derive a quantitative estimate on
validity of the Landau-Zener-Majorana model (\ref{EQ11})-(\ref{EQ14}). The result
is presented in figure \ref{FGErr}. We have used four values of the parameter
$\vare$: $\vare = 0$ (solid line; the egg-crate, i.e. separable lattice), $\vare=1$
(dashed line; the quantum antidot in the terminology of Ref. \cite{Bloch2D2}),
$\vare = 2$ (dotted line), and $\vare =-1$ (dash-dot line; the quantum dot).

From figure \ref{FGErr} one can  see that for the lattice potentials with $V_0\le
0.05$ the  relative error of the finite-dimensional approximation of Zener
tunneling by the Landau-Zener-Majorana models is below $5\%$. For the experimental
lattice of Ref. \cite{PHOT2D}, i.e. the quantum antidot ($\vare =1$), the relative
error is less then $5\%$ for $V_0\le 0.15$. On the other hand, for any particular
lattice potential, for the values of the lattice strength $V_0$ above some
threshold the finite-dimensional approximation will break down, as it does for the
the quantum dot lattice  already for $V_0 =0.15$.

\section{Concluding remarks}
\label{sec4}
We have derived simple models that allow one to describe one-dimensional and more
general two-dimensional Zener tunneling in two-dimensional periodic photonic
structures and calculate the corresponding tunneling probabilities.

We have found that a square two-dimensional photonic lattice allows one to observe
the quasi one-dimensional tunneling, when the Bloch index crosses either the
$\Gamma M$-border or the $XM$-border of the irreducible Brillouin zone away from
the $M$-point, and the specific two-dimensional tunneling, when the Bloch index
passes through the border of the Brillouin zone at the $M$-point (more precisely,
in its neighborhood with radius of the order of $V_0$). We notice that in the first
order in $V_0$ the Bloch oscillations are not affected by  tunneling when the Bloch
index crosses the $\Gamma X$-border of the irreducible Brillouin zone of the
two-dimensional lattice given by Eq. (\ref{lattpot}) of section \ref{sec2}.

In our analytical calculations we have employed a shallow-lattice approximation
which can be related to the experimental setup used in Ref.~\cite{PHOT2D} only
qualitatively, and we do observe at least the qualitative correspondence with the
experiment. There exist two general ways to realize a shallow lattice in the
experimental setup. The first one is to reduce the power of the laser beams
creating the optical lattice potential $I_g(\bm{\xi})$; this requires the condition
$A^2 \ll 1/\varkappa$, which we have adopted in this paper. The second one is to
change the value of the parameter $\varkappa = \gamma n_0(d/\lambda)^2$ so that
$\varkappa \ll 1$. In the latter case, the lattice potential will be different from
that used in our analytic calculations [see Eq.~(\ref{lattpot})], namely it will
have all higher harmonics in the Fourier expansion. Qualitatively this will result
in the following changes. First of all, there will be a substantial tunneling when
crossing the $\Gamma X$-border. Second, the higher harmonics will also affect the
probabilities of all other tunneling cases. However, the latter effect is of a
higher order in the potential amplitude, since the higher harmonics are not
resonant. Therefore, the theory developed here can be applied to this case as well.
The tunneling when crossing the $\Gamma X$-border is quasi one-dimensional, and it
can be treated  similar  as the quasi one-dimensional tunneling cases discussed in
this paper.

\section*{Acknowledgements}

This work was supported by the CNPq-FAPEAL grant in Brazil and the
Australian Research Council in Australia. YK thanks Anton
Desyatnikov for useful discussions of the applicability of the
shallow-lattice approximation to the experimental conditions.


\begin{thebibliography}{99}
\bibitem{note} It is well-know that for a strong linear potential the Bloch-band structure is replaced by the
set of the Wannier-Stark states~\cite{Wan}; see also the discussion in
Ref.~\cite{WS}.
\end{thebibliography}
\end{document}